# scientific reports

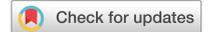

# OPEN  Patents and intellectual property assets as non-fungible tokens; key technologies and challenges

Seyed Mojtaba Hosseini Bamakan[1,4], Nasim Nezhadsistani[2], Omid Bodaghi[3] & Qiang Qu[4,5]✉

With the explosive development of decentralized finance, we witness a phenomenal growth in tokenization of all kinds of assets, including equity, funds, debt, and real estate. By taking advantage of blockchain technology, digital assets are broadly grouped into fungible and non-fungible tokens (NFT). Here non-fungible tokens refer to those with unique and non-substitutable properties. NFT has widely attracted attention, and its protocols, standards, and applications are developing exponentially. It has been successfully applied to digital fantasy artwork, games, collectibles, etc. However, there is a lack of research in utilizing NFT in issues such as Intellectual Property. Applying for a patent and trademark is not only a time-consuming and lengthy process but also costly. NFT has considerable potential in the intellectual property domain. It can promote transparency and liquidity and open the market to innovators who aim to commercialize their inventions efficiently. The main objective of this paper is to examine the requirements of presenting intellectual property assets, specifically patents, as NFTs. Hence, we offer a layered conceptual NFT-based patent framework. Furthermore, a series of open challenges about NFT-based patents and the possible future directions are highlighted. The proposed framework provides fundamental elements and guidance for businesses in taking advantage of NFTs in real-world problems such as grant patents, funding, biotechnology, and so forth.

Distributed ledger technologies (DLTs) such as blockchain are emerging technologies posing a threat to existing business models. Traditionally, most companies used centralized authorities in various aspects of their business, such as financial operations and setting up a trust with their counterparts. By the emergence of blockchain, centralized organizations can be substituted with a decentralized group of resources and actors. The blockchain mechanism was introduced in Bitcoin white paper in 2008, which lets users generate transactions and spend their money without the intervention of banks[1]. Ethereum, which is a second generation of blockchain, was introduced in 2014, allowing developers to run smart contracts on a distributed ledger. With smart contracts, developers and businesses can create financial applications that use cryptocurrencies and other forms of tokens for applications such as decentralized finance (DeFi), crowdfunding, decentralized exchanges, data records keeping, etc.[2]. Recent advances in distributed ledger technology have developed concepts that lead to cost reduction and the simplification of value exchange. Nowadays, by leveraging the advantages of blockchain and taking into account the governance issues, digital assets could be represented as tokens that existed in the blockchain network, which facilitates their transmission and traceability, increases their transparency, and improves their security[3].

In the landscape of blockchain technology, there could be defined two types of tokens, including fungible tokens, in which all the tokens have equal value and non-fungible tokens (NFTs) that feature unique characteristics and are not interchangeable. Actually, non-fungible tokens are digital assets with a unique identifier that is stored on a blockchain[4]. NFT was initially suggested in Ethereum Improvement Proposals (EIP)-721[5], and it was later expanded in EIP-1155[6]. NFTs became one of the most widespread applications of blockchain technology that reached worldwide attention in early 2021. They can be digital representations of real-world objects. NFTs are tradable rights of digital assets (pictures, music, films, and virtual creations) where ownership is recorded in blockchain smart contracts[7].

In particular, fungibility is the ability to exchange one with another of the same kind as an essential currency feature. The non-fungible token is unique and therefore cannot be substituted[8]. Recently, blockchain enthusiasts have indicated significant interest in various types of NFTs. They enthusiastically participate in NFT-related games or trades. CryptoPunks[9], as one of the first NFTs on Ethereum, has developed almost 10,000 collectible punks and helped popularize the ERC-721 Standard. With the gamification of the breeding mechanics, CryptoKitties[10] officially placed NFTs at the forefront of the market in 2017. CryptoKitties is an early blockchain game that enables users to buy, sell, collect, and digital breed cats. Another example is NBA Top Shot[11], an NFT trading platform for digital short films buying and selling NBA events.

[1]Department of Industrial Management, Yazd University, Yazd City, Iran. [2]Department of Electrical and Computer Engineering, Isfahan University of Technology, Isfahan City, Iran. [3]School of Electrical and Computer Engineering, University of Tehran, Tehran City, Iran. [4]Shenzhen Institutes of Advanced Technology, Chinese Academy of Sciences, Shenzhen 518055, China. [5]Huawei Blockchain Lab, Huawei Cloud Tech Co., Ltd., Shenzhen, China. ✉email: qiang@siat.ac.cn

NFTs are developing remarkably and have provided many applications such as artist royalties, in-game assets, educational certificates, etc. However, it is a relatively new concept, and many areas of application need to be explored. Intellectual Property, including patent, trademark, and copyright, is an important area where NFTs can be applied usefully and solve existing problems.

Although NFTs have had many applications so far, it rarely has been used to solve real-world problems. In fact, an NFT is an exciting concept about Intellectual Property (IP). Applying for a patent and trademark is a time-consuming and lengthy process, but it is also costly. That is, registering a copyright or trademark may take months, while securing a patent can take years. On the contrary, with the help of unique features of NFT technology, it is possible to accelerate this process with considerable confidence and assurance about protecting the ownership of an IP. NFTs can offer IP protection while an applicant waits for the government to grant his/her more formal protection. It is cause for excitement that people who believe NFTs and Blockchain would make buying and selling patents easier, offering new opportunities for companies, universities, and inventors to make money off their innovations[12]. Patent holders will benefit from such innovation. It would give them the ability to 'tokenize' their patents. Because every transaction would be logged on a blockchain, it will be much easier to trace patent ownership changes. However, NFT would also facilitate the revenue generation of patents by democratizing patent licensing via NFT. NFTs support the intellectual property market by embedding automatic royalty collecting methods inside inventors' works, providing them with financial benefits anytime their innovation is licensed. For example, each inventor's patent would be minted as an NFT, and these NFTs would be joined together to form a commercial IP portfolio and minted as a compounded NFT. Each investor would automatically get their fair share of royalties whenever the licensing revenue is generated without tracking them down.

The authors in[13], an overview of NFTs' applications in different aspects such as gambling, games, and collectibles has been discussed. In addition[4], provides a prototype for an event-tracking application based on Ethereum smart contract, and NFT as a solution for art and real estate auction systems is described in[14]. However, these studies have not discussed existing standards or a generalized architecture, enabling NFTs to be applied in diverse applications. For example, the authors in[15] provide two general design patterns for creating and trading NFTs and discuss existing token standards for NFT. However, the proposed designs are limited to Ethereum, and other blockchains are not considered[16]. Moreover, different technologies for each step of the proposed procedure are not discussed. In[8], the authors provide a conceptual framework for token designing and managing and discuss five views: token view, wallet view, transaction view, user interface view, and protocol view. However, no research provides a generalized conceptual framework for generating, recording, and tracing NFT based-IP, in blockchain network.

Even with the clear benefits that NFT-backed patents offer, there are a number of impediments to actually achieving such a system. For example, convincing patent owners to put current ownership records for their patents into NFTs poses an initial obstacle. Because there is no reliable framework for NFT-based patents, this paper provides a conceptual framework for presenting NFT-based patents with a comprehensive discussion on many aspects, ranging from the background, model components, token standards to application domains and research challenges. The main objective of this paper is to provide a layered conceptual NFT-based patent framework that can be used to register patents in a decentralized, tamper-proof, and trustworthy peer-to-peer network to trade and exchange them in the worldwide market. The main contributions of this paper are highlighted as follows:

- Providing a comprehensive overview on tokenization of IP assets to create unique digital tokens.
- Discussing the components of a distributed and trustworthy framework for minting NFT-based patents.
- Highlighting a series of open challenges of NFT-based patents and enlightening the possible future trends.

The rest of the paper is structured as follows: "Background" section describes the Background of NFTs, Non-Fungible Token Standards. The NFT-based patent framework is described in "NFT-based patent framework" section. The Discussion and challenges are presented in "Discussion" section. Lastly, conclusions are given in "Conclusion" section.

## Background

Colored Coins could be considered the first steps toward NFTs designed on the top of the Bitcoin network. Bitcoins are fungible, but it is possible to mark them to be distinguishable from the other bitcoins. These marked coins have special properties representing real-world assets like cars and stocks, and owners can prove their ownership of physical assets through the colored coins. By utilizing Colored Coins, users can transfer their marked coins' ownership like a usual transaction and benefit from Bitcoin's decentralized network[17]. Colored Coins had limited functionality due to the Bitcoin script limitations. Pepe is a green frog meme originated by Matt Furie that; users define tokens for Pepes and trade them through the Counterparty platform. Then, the tokens that were created by the picture of Pepes are decided if they are rare enough. Rare Pepe allows users to preserve scarcity, manage the ownership, and transfer their purchased Pepes.

In 2017, Larva Labs developed the first Ethereum-based NFT named CryptoPunks. It contains 10,000 unique human-like characters generated randomly. The official ownership of each character is stored in the Ethereum smart contract, and owners would trade characters. CryptoPunks project inspired CryptoKitties project.

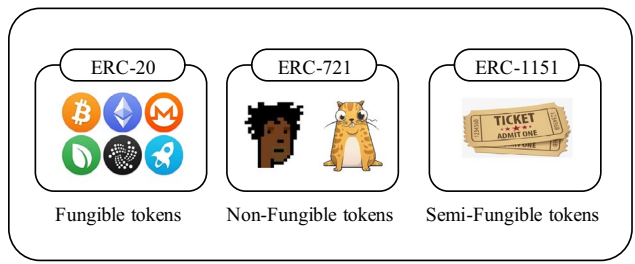

**Figure 1.** From right to left (fungible, non-fungible, semi-fungible).

CryptoKitties attracts attention to NFT, and it is a pioneer in blockchain games and NFTs that launched in late 2017. CryptoKitties is a blockchain-based virtual game, and users collect and trade characters with unique features that shape kitties. This game was developed in Ethereum smart contract, and it pioneered the ERC-721 token, which was the first standard token in the Ethereum blockchain for NFTs. After the 2017 hype in NFTs, many projects started in this context. Due to increased attention to NFTs' use-cases and growing market cap, different blockchains like EOS, Algorand, and Tezos started to support NFTs, and various marketplaces like SuperRare and Rarible, and OpenSea are developed to help users to trade NFTs. As mentioned, in general, assets are categorized into two main classes, fungible and non-fungible assets. Fungible assets are the ones that another similar asset can replace. Fungible items could have two main characteristics: replicability and divisibility.

Currency is a fungible item because a ten-dollar bill can be exchanged for another ten-dollar bill or divided into ten one-dollar bills. Despite fungible items, non-fungible items are unique and distinguishable. They cannot be divided or exchanged by another identical item. The first tweet on Twitter is a non-fungible item with mentioned characteristics. Another tweet cannot replace it, and it is unique and not divisible. NFT is a non-fungible cryptographic asset that is declared in a standard token format and has a unique set of attributes. Due to transparency, proof of ownership, and traceable transactions in the blockchain network, NFTs are created using blockchain technology.

Blockchain-based NFTs help enthusiasts create NFTs in the standard token format in blockchain, transfer the ownership of their NFTs to a buyer, assure uniqueness of NFTs, and manage NFTs completely. In addition, there are semi-fungible tokens that have characteristics of both fungible and non-fungible tokens. Semi-fungible tokens are fungible in the same class or specific time and non-fungible in other classes or different times. A plane ticket can be considered a semi-fungible token because a charter ticket can be exchanged by another charter ticket but cannot be exchanged by a first-class ticket. The concept of semi-fungible tokens plays the main role in blockchain-based games and reduces NFTs overhead. In Fig. 1, we illustrate fungible, non-fungible, and semi-fungible tokens. The main properties of NFTs are described as follows[15]:

*Ownership*: Because of the blockchain layer, the owner of NFT can easily prove the right of possession by his/her keys. Other nodes can verify the user's ownership publicly.

*Transferable*: Users can freely transfer owned NFTs ownership to others on dedicated markets.
*Transparency*: By using blockchain, all transactions are transparent, and every node in the network can confirm and trace the trades.
*Fraud Prevention*: Fraud is one of the key problems in trading assets; hence, using NFTs ensures buyers buy a non-counterfeit item.
*Immutability*: Metadata, token ID, and history of transactions of NFTs are recorded in a distributed ledger, and it is impossible to change the information of the purchased NFTs.

**Non-fungible standards.** Ethereum blockchain was pioneered in implementing NFTs. ERC-721 token was the first standard token accepted in the Ethereum network. With the increase in popularity of the NFTs, developers started developing and enhancing NFTs standards in different blockchains like EOS, Algorand, and Tezos. This section provides a review of implemented NFTs standards on the mentioned blockchains.

*Ethereum.* ERC-721 was the first Standard for NFTs developed in Ethereum, a free and open-source standard. ERC-721 is an interface that a smart contract should implement to have the ability to transfer and manage NFTs. Each ERC-721 token has unique properties and a different Token Id. ERC-721 tokens include the owner's information, a list of approved addresses, a transfer function that implements transferring tokens from owner to buyer, and other useful functions[5].

In ERC-721, smart contracts can group tokens with the same configuration, and each token has different properties, so ERC-721 does not support fungible tokens. However, ERC-1155 is another standard on Ethereum developed by Enjin and has richer functionalities than ERC-721 that supports fungible, non-fungible, and semi-fungible tokens. In ERC-1155, IDs define the class of assets. So different IDs have a different class of assets, and each ID may contain different assets of the same class. Using ERC-1155, a user can transfer different types of tokens in a single transaction and mix multiple fungible and non-fungible types of tokens in a single smart contract[6]. ERC-721 and ERC-1155 both support operators in which the owner can let the operator originate transferring of the token.

| NFT standards | Support FTs | Support NFTs | Support batch transferring | Support operator | Fractionalized NFTs |
|---|---|---|---|---|---|
| ERC-721 | No | Yes | No | Yes | No |
| ERC-1155 | Yes | Yes | Yes | Yes | No |
| dGoods | Yes | Yes | Yes | No | No |
| Algorand | Yes | Yes | Yes | Yes | Yes |
| Tezos | Yes | Yes | Yes | Yes | Yes |
| Flow | Yes | Yes | Yes | Yes | No |

**Table 1.** Comparing NFT standards.

*EOSIO.* EOSIO is an open-source blockchain platform released in 2018 and claims to eliminate transaction fees and increase transaction throughput. EOSIO differs from Ethereum in the wallet creation algorithm and procedure of handling transactions. dGood is a free standard developed in the EOS blockchain for assets, and it focuses on large-scale use cases. It supports a hierarchical naming structure in smart contracts. Each contract has a unique symbol and a list of categories, and each category contains a list of token names. Therefore, a single contract in dGoods could contain many tokens, which causes efficiency in transferring a group of tokens. Using this hierarchy, dGoods supports fungible, non-fungible, and semi-fungible tokens. It also supports batch transferring, where the owner can transfer many tokens in one operation[18].

*Algorand.* Algorand is a new high-performance public blockchain launched in 2019. It provides scalability while maintaining security and decentralization. It supports smart contracts and tokens for representing assets[19]. Algorand defines Algorand Standard Assets (ASA) concept to create and manage assets in the Algorand blockchain. Using ASA, users are able to define fungible and non-fungible tokens. In Algorand, users can create NFTs or FTs without writing smart contracts, and they should run just a single transaction in the Algorand blockchain. Each transaction contains some mutable and immutable properties[20].

Each account in Algorand can create up to 1000 assets, and for every asset, an account creates or receives, the minimum balance of the account increases by 0.1 Algos. Also, Algorand supports fractional NFTs by splitting an NFT into a group of divided FTs or NFTs, and each part can be exchanged dependently[21]. Algorand uses a Clawback Address that operates like an operator in ERC-1155, and it is allowed to transfer tokens of an owner who has permitted the operator.

*Tezos.* Tezos is another decentralized open-source blockchain. Tezos supports the meta-consensus concept. In addition to using a consensus protocol on the ledger's state like Bitcoin and Ethereum, It also attempts to reach a consensus about how nodes and the protocol should change or upgrade[22]. FA2 (TZIP-12) is a standard for a unified token contract interface in the Tezos blockchain. FA2 supports different token types like fungible, non-fungible, and fractionalized NFT contracts. In Tezos, tokens are identified with a token contract address and token ID pair. Also, Tezos supports batch token transferring, which reduces the cost of transferring multiple tokens.

*Flow.* Flow was developed by Dapper Labs to remove the scalability limitation of the Ethereum blockchain. Flow is a fast and decentralized blockchain that focuses on games and digital collectibles. It improves throughput and scalability without sharding due to its architecture. Flow supports smart contracts using Cadence, which is a resource-oriented programming language. NFTs can be described as a resource with a unique id in Cadence. Resources have important rules for ownership management; that is, resources have just one owner and cannot be copied or lost. These features assure the NFT owner. NFTs' metadata, including images and documents, can be stored off-chain or on-chain in Flow. In addition, Flow defines a Collection concept, in which each collection is an NFT resource that can include a list of resources. It is a dictionary that the key is resource id, and the value is corresponding NFT.

The collection concept provides batch transferring of NFTs. Besides, users can define an NFT for an FT. For instance, in CryptoKitties, a unique cat as an NFT can own a unique hat (another NFT). Flow uses Cadence's second layer of access control to allow some operators to access some fields of the NFT[23]. In Table 1, we provide a comparison between explained standards. They are compared in support of fungible-tokens, non-fungible tokens, batch transferring that owner can transform multiple tokens in one operation, operator support in which the owner can approve an operator to originate token transfer, and fractionalized NFTs that an NFT can divide to different tokens and each exchange dependently.

## NFT-based patent framework

In this section, we propose a framework for presenting NFT-based patents. We describe details of the proposed distributed and trustworthy framework for minting NFT-based patents, as shown in Fig. 2. The proposed framework includes five main layers: Storage Layer, Authentication Layer, Verification Layer, Blockchain Layer, and Application Layer. Details of each layer and the general concepts are presented as follows.

**Storage layer.** The continuous rise of the data in blockchain technology is moving various information systems towards the use of decentralized storage networks. Decentralized storage networks were created to provide more benefits to the technological world[24]. Some of the benefits of using decentralized storage systems are explained: (1) Cost savings are achieved by making optimal use of current storage. (2) Multiple copies are kept on various nodes, avoiding bottlenecks on central servers and speeding up downloads. This foundation layer implicitly provides the infrastructure required for the storage. The items on NFT platforms have unique characteristics that must be included for identification.

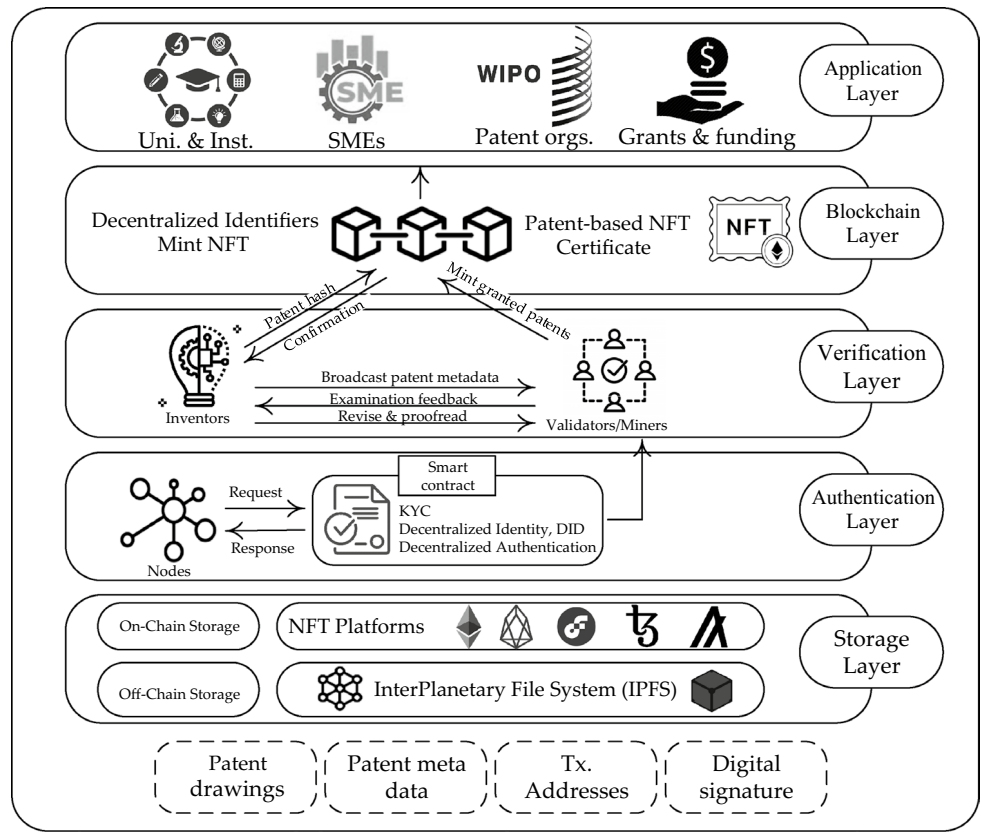

**Figure 2.** Proposed Architecture for Presenting NFT-based patent.

Non-fungible token metadata provides information that describes a particular token ID. NFT metadata is either represented on the On-chain or Off-chain. *On-chain* means direct incorporation of the metadata into the NFT's smart contract, which represents the tokens. On the other hand, off-chain storage means hosting the metadata separately[25].

Blockchains provide decentralization but are expensive for data storage and never allow data to be removed. For example, because of the Ethereum blockchain's current storage limits and high maintenance costs, many projects' metadata is maintained off-chain. Developers utilize the ERC721 Standard, which features a method known as tokenURI. This method is implemented to let applications know the location of the metadata for a specific item. Currently, there are three solutions for off-chain storage, including InterPlanetary File System (IPFS), Pinata, and Filecoin.

*IPFS.* InterPlanetary File System (IPFS) is a peer-to-peer hypermedia protocol for decentralized media content storage. Because of the high cost of storing media files related to NFTS on Blockchain, IPFS can be the most affordable and efficient solution. IPFS combines multiple technologies inspired by Gita and BitTorrent, such as Block Exchange System, Distributed Hash Tables (DHT), and Version Control System[26]. On a peer-to-peer network, DHT is used to coordinate and maintain metadata.

In other words, the hash values must be mapped to the objects they represent. An IPFS generates a hash value that starts with the prefix $Q_m$ and acts as a reference to a specific item when storing an object like a file. Objects larger than 256 KB are divided into smaller blocks up to 256 KB. Then a hash tree is used to interconnect all the blocks that are a part of the same object. IPFS uses Kamdelia DHT. The Block Exchange System, or BitSwap, is a BitTorrent-inspired system that is used to exchange blocks. It is possible to use asymmetric encryption to prevent unauthorized access to stored content on IPFS[27].

*Pinata.* Pinata is a popular platform for managing and uploading files on IPFS. It provides secure and verifiable files for NFTs. Most data is stored *off-chain* by most NFTs, where a URL of the data is pointed to the NFT on the blockchain. The main problem here is that some information in the URL can change.

This indicates that an NFT supposed to describe a certain patent can be changed without anyone knowing. This defeats the purpose of the NFT in the first place. This is where Pinata comes in handy. Pinata uses the IPFS to create content-addressable hashes of data, also known as Content-Identifiers (CIDs). These CIDs serve as both a way of retrieving data and a means to ensure data validity. Those looking to retrieve data simply ask the IPFS network for the data associated with a certain CID, and if any node on the network contains that data, it will be returned to the requester. The data is automatically rehashed on the requester's computer when the requester retrieves it to make sure that the data matches back up with the original CID they asked for. This process ensures the data that's received is exactly what was asked for; if a malicious node attempts to send fake data, the resulting CID on the requester's end will be different, alerting the requester that they're receiving incorrect data[28].

*Filecoin.* Another decentralized storage network is Filecoin. It is built on top of IPFS and is designed to store the most important data, such as media files. Truffle Suite has also launched NFT Development Template with Filecoin Box. NFT.Storage (Free Decentralized Storage for NFTs)[29] allows users to easily and securely store their NFT content and metadata using IPFS and Filecoin. NFT.Storage is a service backed by Protocol Labs and Pinata specifically for storing NFT data. Through content addressing and decentralized storage, NFT.Storage allows developers to protect their NFT assets and associated metadata, ensuring that all NFTs follow best practices to stay accessible for the long term. NFT.Storage makes it completely frictionless to mint NFTs following best practices through resilient persistence on IPFS and Filecoin. NFT.Storage allows developers to quickly, safely, and for free store NFT data on decentralized networks. Anyone can leverage the power of IPFS and Filecoin to ensure the persistence of their NFTs. The details of this system are stated as follows[30]:

Content addressing. Once users upload data on NFT.Storage, They receive a CID, which is an IPFS hash of the content. CIDs are the data's unique fingerprints, universal addresses that can be used to refer to it regardless of how or where it is stored. Using CIDs to reference NFT data avoids problems such as weak links and "rug pulls" since CIDs are generated from the content itself.

Provable storage. NFT.Storage uses Filecoin for long-term decentralized data storage. Filecoin uses cryptographic proofs to assure the NFT data's durability and persistence over time.

Resilient retrieval. This data stored via IPFS and Filecoin can be fetched directly in the browser via any public IPFS.

**Authentication Layer.** The second layer is the authentication layer, which we briefly highlight its functions in this section. The Decentralized Identity (DID) approach assists users in collecting credentials from a variety of issuers, such as the government, educational institutions, or employers, and saving them in a digital wallet. The verifier then uses these credentials to verify a person's validity by using a blockchain-based ledger to follow the "identity and access management (IAM)" process. Therefore, DID allows users to be in control of their identity. A lack of NFT verifiability also causes intellectual property and copyright infringements; of course, the chain of custody may be traced back to the creator's public address to check whether a similar patent is filed using that address. However, there is no quick and foolproof way to check an NFTs creator's legitimacy. Without such verification built into the NFT, an NFT proves ownership only over that NFT itself and nothing more.

Self-sovereign identity (SSI)[31] is a solution to this problem. SSI is a new series of standards that will guide a new identity architecture for the Internet. With a focus on privacy, security interoperability, SSI applications use public-key cryptography with public blockchains to generate persistent identities for people with private and selective information disclosure. Blockchain technology offers a solution to establish trust and transparency and provide a secure and publicly verifiable KYC (Know Your Customer). The blockchain architecture allows you to collect information from various service providers into a single cryptographically secure and unchanging database that does not need a third party to verify the authenticity of the information.

The proposed platform generates patents-related smart contracts acting as a program that runs on the blockchain to receive and send transactions. They are unalterable privately identifying clients with a thorough KYC process. After KYC approval, then mint an NFT on the blockchain as a certificate of verification[32]. This article uses a decentralized authentication solution at this layer for authentication. This solution has been used for various applications in the field of the blockchain (exp: smart city, Internet of Things, etc.[33, 34], but we use it here for the proposed framework (patent as NFTs). Details of this solution will be presented in the following.

*Decentralized authentication.* This section presents the authentication layer similar[35] to build validated communication in a secure and decentralized manner via blockchain technology. As shown in Fig. 3, the authentication protocol comprises two processes, including registration and login.

Registration. In the registration process of a suggested authentication protocol, we first initialize a user's public key as their identity key (UserName). Then, we upload this identity key on a blockchain, in which transactions can be verified later by other users. Finally, the user generates an identity transaction.

Login. After registration, a user logs in to the system. The login process is described as follows:

1. The user commits identity information and imports their secret key into the service application to log in.
2. A user who needs to log in sends a login request to the network's service provider.
3. The service provider analyzes the login request, extracts the hash, queries the blockchain, and obtains identity information from an identity list (identity transactions).

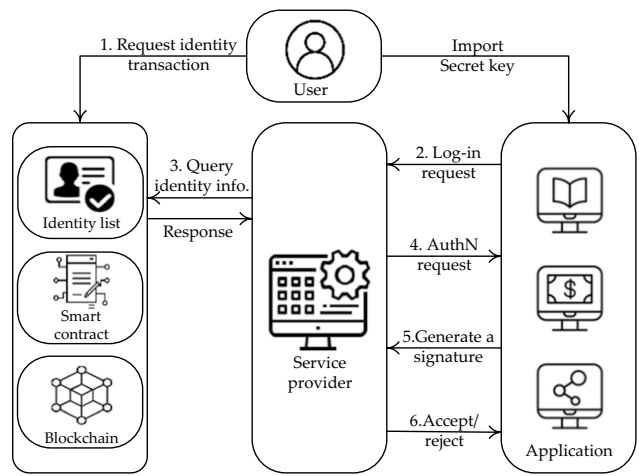

**Figure 3.** Decentralized Authentication.

4. The service provider responds with an authentication request when the above process is completed. A timestamp (to avoid a replay attack), the user's UserName, and a signature are all included in the authentication request.
5. The user creates a signature with five parameters: timestamp, UserName, and PK, as well as the UserName and PK of the service provider. The user authentication credential is used as the signature.
6. The service provider verifies the received information, and if the received information is valid, the authentication succeeds; otherwise, the authentication fails, and the user's login is denied.

The World Intellectual Property Organization (WIPO) and multiple target patent offices in various nations or regions should assess a patent application, resulting in inefficiency, high costs, and uncertainty. This study presented a conceptual NFT-based patent framework for issuing, validating, and sharing patent certificates. The platform aims to support counterfeit protection as well as secure access and management of certificates according to the needs of learners, companies, education institutions, and certification authorities.

Here, the certification authority (CA) is used to authenticate patent offices. The procedure will first validate a patent if it is provided with a digital certificate that meets the X.509 standard. Certificate authorities are introduced into the system to authenticate both the nodes and clients connected to the blockchain network.

**Verification layer.** In permissioned blockchains, just identified nodes can read and write in the distributed ledger. Nodes can act in different roles and have various permissions. Therefore, a distributed system can be designed to be the identified nodes for patent granting offices. Here the system is described conceptually at a high level. Figure 4 illustrates the sequence diagram of this layer. This layer includes four levels as below:

*Digitalization.* For a patent to publish as an NFT in the blockchain, it must have a digitalized format. This level is the "filling step" in traditional patent registering. An application could be designed in the application layer to allow users to enter different patent information online.

*Recording.* Patents provide valuable information and would bring financial benefits for their owner. If they are publicly published in a blockchain network, miners may refuse the patent and take the innovation for themselves. At least it can weaken consensus reliability and encourage miners to misbehave. The inventor should record his innovation privately first using proof of existence to prevent this. The inventor generates the hash of the patent document and records it in the blockchain. As soon as it is recorded in the blockchain, the timestamp and the hash are available for others publicly. Then, the inventor can prove the existence of the patent document whenever it is needed.

Furthermore, using methods like Decision Thinking[36], an inventor can record each phase of patent development separately. In each stage, a user generates the hash of the finished part and publishes the hash regarding the last part's hash. Finally, they have a coupled series of hashes that indicate patent development, and they can prove the existence of each phase using the original related documents. This level should be done to prevent others from abusing the patent and taking it for themselves. The inventor can make sure that their patent document is recorded confidentially and immutably[37].

Different hash algorithms exist with different architecture, time complexity, and security considerations. Hash functions should satisfy two main requirements: Pre-Image Resistance: This means that it should be computationally hard to find the input of a hash function while the output and the hash algorithm are known publicly. Collision Resistance: This means that it is computationally hard to find two arbitrary inputs, x, and y, that have the same hash output. These requirements are vital for recording patents. First, the hash function should be Pre-Image Resistance to make it impossible for others to calculate the patent documentation. Otherwise, everybody can read the patent, even before its official publication. Second, the hash function should satisfy Collision Resistance to preclude users from changing their document after recording. Otherwise, users can upload another document, and after a while, they can replace it with another one.

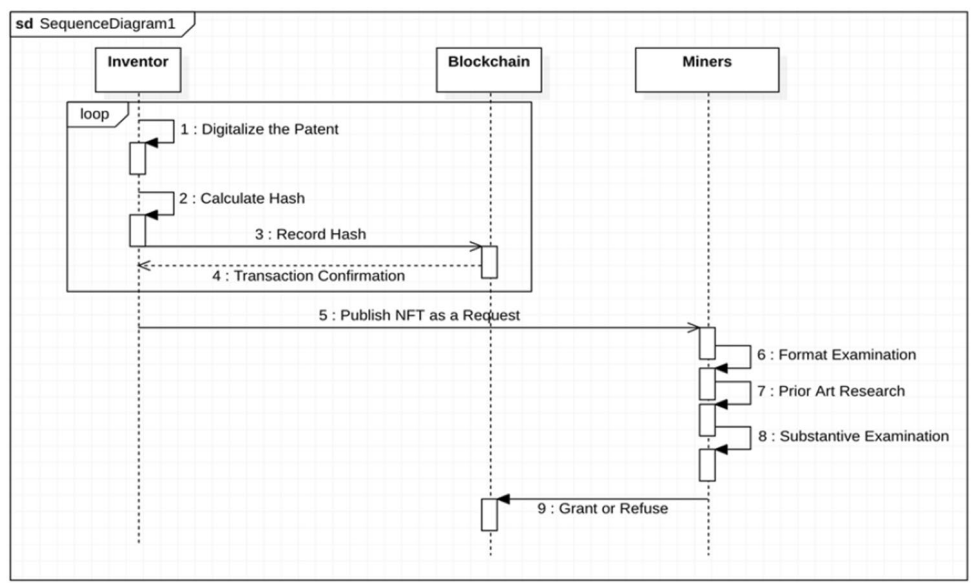

**Figure 4.** Verification layer sequence diagram.

There are various hash algorithms, and MD and SHA families are the most useful algorithms. According to[38], Collisions have been found for MD2, MD4, MD5, SHA-0, and SHA-1 hash functions. Hence, they cannot be a good choice for recording patents. SHA2 hash algorithm is secure, and no collision has been found. Although SHA2 is noticeably slower than prior hash algorithms, the recording phase is not highly time-sensitive. So, it is a better choice and provides excellent security for users.

*Validating.* In this phase, the inventors first create NFT for their patents and publish it to the miners/validators. Miners are some identified nodes that validate NFTs to record in the blockchain. Due to the specialization of the patent validation, miners cannot be inexpert public persons. In addition, patent offices are not too many to make the network fully decentralized. Therefore, the miners can be related specialist persons that are certified by the patent offices. They should receive a digital certificate from patent offices that show their eligibility to referee a patent.

*Digital certificate.* Digital certificates are digital credentials used to verify networked entities' online identities. They usually include a public key as well as the owner's identification. They are issued by Certification Authorities (CAs), who must verify the certificate holder's identity. Certificates contain cryptographic keys for signing, encryption, and decryption. X.509 is a standard that defines the format of public-key certificates and is signed by a certificate authority. X.509 standard has multiple fields, and its structure is shown in Fig. 5. Version: This field indicated the version of the X.509 standard. X.509 contains multiple versions, and each version has a different structure. According to the CA, validators can choose their desired version. Serial Number: It is used to distinguish a certificate from other certificates. Thus, each certificate has a unique serial number. Signature Algorithm Identifier: This field indicates the cryptographic encryption algorithm used by a certificate authority. Issuer Name: This field indicates the issuer's name, which is generally certificate authority. Validity Period: Each certificate is valid for a defined period, defined as the Validity Period. This limited period partly protects certificates against exposing CA's private key. Subject Name: Name of the requester. In our proposed framework, it is the validator's name. Subject Public Key Info: Shows the CA's or organization's public key that issued the certificate. These fields are identical among all versions of the X.509 standard[39].

*Certificate authority.* A Certificate Authority (CA) issues digital certificates. CAs encrypt the certificate with their private key, which is not public, and others can decrypt the certificates containing the CA's public key.

Here, the patent office creates a certificate for requested patent referees. The patent office writes the information of the validator in their certificate and encrypts it with the patent offices' private key. The validator can use the certificate to assure others about their eligibility. Other nodes can check the requesting node's information by decrypting the certificate using the public key of the patent office. Therefore, persons can join the network's miners/validators using their credentials. In this phase, miners perform Formal Examinations, Prior Art Research, and Substantive Examinations and vote to grant or refuse the patent.

Miners perform a consensus about the patent and record the patent in the blockchain. After that, the NFT is recorded in the blockchain with corresponding comments in granting or needing reformations. If the miners detect the NFT as a malicious request, they do not record it in the blockchain.

**Figure 5.** X.509 certificates structure.

| Classification | Consensus algorithms |
| --- | --- |
| Voting-based consensus | Practical Byzantine Fault Tolerance (PBFT)[44], Yet another consensus (YAC) protocol[45], BFT-SMaRT protocol[46], Ripple[47], Raft[48], Delegated Proof of Stake (DPoS)[49], Paxos[50], Clique / Aura protocols[51] |
| Chain-based consensus | Proof of Work (PoW)[1], Proof of Stake (PoS)[52], Proof of Elapsed Time (PoET)[53] Proof of Weight (PoWeight)[54], Proof of Burn (PoB)[55], Proof of capacity (PoC)[56], Proof of Activity (PoA)[57], Proof of importance (PoI)[58] |
| DAG-based consensus | IoTA (Tangle )[59], HashGraph[60], Nano[61] |

**Table 2.** Consensus algorithms in blockchain networks.

**Blockchain layer.**  This layer plays as a middleware between the Verification Layer and Application Layer in the patents as NFTs architecture. The main purpose of the blockchain layer in the proposed architecture is to provide IP management. We find that transitioning to a blockchain-based patent as a NFTs records system enables many previously suggested improvements to current patent systems in a flexible, scalable, and transparent manner.

On the other hand, we can use multiple blockchain platforms, including Ethereum, EOS, Flow, and Tezos. Blockchain Systems can be mainly classified into two major types: Permissionless (public) and Permissioned (private) Blockchains based on their consensus mechanism. In a public blockchain, any node can participate in the peer-to-peer network, where the blockchain is fully decentralized. A node can leave the network without any consent from the other nodes in the network.

Bitcoin is one of the most popular examples that fall under the public and permissionless blockchain. Proof of Work (POW), Proof-of-Stake (POS), and directed acyclic graph (DAG) are some examples of consensus algorithms in permissionless blockchains. Bitcoin and Ethereum, two famous and trustable blockchain networks, use the PoW consensus mechanism. Blockchain platforms like Cardano and EOS adopt the PoS consensus[40].

Nodes require specific access or permission to get network authentication in a private blockchain. Hyperledger is among the most popular private blockchains, which allow only permissioned members to join the network after authentication. This provides security to a group of entities that do not completely trust one another but wants to achieve a common objective such as exchanging information. All entities of a permissioned blockchain network can use Byzantine-fault-tolerant (BFT) consensus. The Fabric has a membership identity service that manages user IDs and verifies network participants.

Therefore, members are aware of each other's identity while maintaining privacy and secrecy because they are unaware of each other's activities[41]. Due to their more secure nature, private blockchains have sparked a large interest in banking and financial organizations, believing that these platforms can disrupt current centralized systems. Hyperledger, Quorum, Corda, EOS are some examples of permissioned blockchains[42].

Reaching consensus in a distributed environment is a challenge. Blockchain is a decentralized network with no central node to observe and check all transactions. Thus, there is a need to design protocols that indicate all transactions are valid. So, the consensus algorithms are considered as the core of each blockchain[43]. In distributed systems, the consensus has become a problem in which all network members (nodes) agree on accept or reject of a block. When all network members accept the new block, it can append to the previous block.

As mentioned, the main concern in the blockchains is how to reach consensus among network members. A wide range of consensus algorithms has been designed in which each of them has its own pros and cons[42]. Blockchain consensus algorithms are mainly classified into three groups shown in Table 2. As the first group, proof-based consensus algorithms require the nodes joining the verifying network to demonstrate their qualification to do the appending task. The second group is voting-based consensus that requires validators in the network to share their results of validating a new block or transaction before making the final decision. The third group is DAG-based consensus, a new class of consensus algorithms. These algorithms allow several different blocks to be published and recorded simultaneously on the network.

The proposed patent as the NFTs platform that builds blockchain intellectual property empowers the entire patent ecosystem. It is a solution that removes barriers by addressing fundamental issues within the traditional patent ecosystem. Blockchain can efficiently handle patents and trademarks by effectively reducing approval wait time and other required resources. The user entities involved in Intellectual Property management are Creators, Patent Consumers, and Copyright Managing Entities. Users with ownership of the original data are the patent creators, e.g., inventors, writers, and researchers. Patent Consumers are the users who are willing to consume the content and support the creator's work. On the other hand, Users responsible for protecting the creators' Intellectual Property are the copyright management entities, e.g., lawyers. The patents as NFTs solution for IP management in blockchain layer works by implementing the following steps[62]:

*Creators sign up to the platform.* Creators need to sign up on the blockchain platform to patent their creative work. The identity information will be required while signing up.

*Creators upload IP on the blockchain network.* Now, add an intellectual property for which the patent application is required. The creator will upload the information related to IP and the data on the blockchain network. Blockchain ensures traceability and auditability to prevent data from duplicity and manipulation. The patent becomes visible to all network members once it is uploaded to the blockchain.

*Consumers generate request to use the content.* Consumers who want to access the content must first register on the blockchain network. After Signing up, consumers can ask creators to grant access to the patented content. Before the patent owner authorizes the request, a Smart Contract is created to allow customers to access information such as the owner's data. Furthermore, consumers are required to pay fees in either fiat money or unique tokens in order to use the creator's original information. When the creator approves the request, an NDA (Non-Disclosure Agreement) is produced and signed by both parties. Blockchain manages the agreement and guarantees that all parties agree to the terms and conditions filed.

*Patent management entities leverage blockchain to protect copyrights and solve related disputes.* Blockchain assists the patent management entities in resolving a variety of disputes that may include: sharing confidential information, establishing proof of authorship, transferring IP rights, and making defensive publications, etc. Suppose a person used an Invention from a patent for his company without the inventor's consent. The inventor can report it to the patent office and claim that he is the owner of that invention.

**Application layer.** The patent Platform Global Marketplace technology would allow many enterprises, governments, universities, and Small and medium-sized enterprises (SMEs) worldwide to tokenize patents as NFTs to create an infrastructure for storing patent records on a blockchain-based network and developing a decentralized marketplace in which patent holders would easily sell or otherwise monetize their patents. The NFTs-based patent can use smart contracts to determine a set price for a license or purchase.

Any buyer satisfied with the conditions can pay and immediately unlock the rights to the patent without either party ever having to interact directly. While patents are currently regulated jurisdictionally around the world, a blockchain-based patent marketplace using NFTs can reduce the geographical barriers between patent systems using as simple a tool as a search query. The ease of access to patents globally can help aspiring inventors accelerate the innovative process by building upon others' patented inventions through licenses. There are a wide variety of use cases for patent NFTs such as SMEs, Patent Organization, Grant & Funding, and fundraising/transferring information relating to patents. These applications keep growing as time progresses, and we are constantly finding new ways to utilize these tokens. Some of the most commonly used applications can be seen as follows.

*SMEs.* The aim is to move intellectual property assets onto a digital, centralized, and secure blockchain network, enabling easier commercialization of patents, especially for small or medium enterprises (SMEs). Smart contracts can be attached to NFTs so terms of use and ownership can be outlined and agreed upon without incurring as many legal fees as traditional IP transfers. This is believed to help SMEs secure funding, as they could more easily leverage the previously undisclosed value of their patent portfolios[63].

*Transfer ownership of patents.* NFTs can be used to transfer ownership of patents. The blockchain can be used to keep track of patent owners, and tokens would include self-executing contracts that transfer the legal rights associated with patents when the tokens are transferred. A partnership between IBM and IPwe has spearheaded the use of NFTs to secure patent ownership. These two companies have teamed together to build the infrastructure for an NFT-based patent marketplace.

## Discussion

There are exciting proposals in the legal and economic literature that suggest seemingly straightforward solutions to many of the issues plaguing current patent systems. However, most solutions would constitute major administrative disruptions and place significant and continuous financial burdens on patent offices or their users. An NFT-based patents system not only makes many of these ideas administratively feasible but can also be examined in a step-wise, scalable, and very public manner.

Furthermore, NFT-based patents may facilitate reliable information sharing among offices and patentees worldwide, reducing the burden on examiners and perhaps even accelerating harmonization efforts. NFT-based patents also have additional transparency and archival attributes baked in. A patent should be a privilege bestowed on those who take resource-intensive risks to explore the frontier of technological capabilities. As a reward for their achievements, full transparency of these rewards is much public interest. It is a society that pays for administrative and economic inefficiencies that exist in today's systems. NFT-based patents can enhance this transparency. From an organizational perspective, an NFT-based patent can remove current bottlenecks in patent processes by making these processes more efficient, rapid, and convenient for applicants without compromising the quality of granted patents.

The proposed framework encounters some challenges that should be solved to reach a developed patent verification platform. First, technical problems are discussed. The consensus method that is used in the verification layer is not addressed in detail. Due to the permissioned structure of miners in the NFT-based patents, consensus algorithms like PBFT, Federated Consensus, and Round Robin Consensus are designed for permissioned blockchains can be applied. Also, miners/validators spend some time validating the patents; hence a protocol should be designed to profit them. Some challenges like proving the miners' time and effort, the price that inventors should pay to miners, and other economic trade-offs should be considered.

Different NFT standards were discussed. If various patent services use NFT standards, there will be some cross-platform problems. For instance, transferring an NFT from Ethereum blockchain (ERC-721 token) to EOS blockchain is not a forward and straight work and needs some considerations. Also, people usually trade NFTs in marketplaces such as Rarible and OpenSea. These marketplaces are centralized and may prompt some challenges because of their centralized nature. Besides, there exist some other types of challenges. For example, the novelty of NFT-based patents and blockchain services.

Blockchain-based patent service has not been tested before. The patent registration procedure and concepts of the Patent as NFT system may be ambiguous for people who still prefer conventional centralized patent systems over decentralized ones. It should be noted that there are some problems in the mining part. Miners should receive certificates from the accepted organizations. Determining these organizations and how they accept referees as validators need more consideration. Some types of inventions in some countries are prohibited, and inventors cannot register them. In NFT-based patents, inventors can register their patents publicly, and maybe some collisions occur between inventors and the government. There exist some misunderstandings about NFT's ownership rights. It is not clear that when a person buys an NFT, which rights are given to them exactly; for instance, they have property rights or have moral rights, too.

## Conclusion

Blockchain technology provides strong timestamping, the potential for smart contracts, proof-of-existence. It enables creating a transparent, distributed, cost-effective, and resilient environment that is open to all and where each transaction is auditable. On the other hand, blockchain is a definite boon to the IP industry, benefitting patent owners. When blockchain technology's intrinsic characteristics are applied to the IP domain, it helps copyrights. This paper provided a conceptual framework for presenting an NFT-based patent with a comprehensive discussion of many aspects: background, model components, token standards to application areas, and research challenges. The proposed framework includes five main layers: Storage Layer, Authentication Layer, Verification Layer, Blockchain Layer, and Application. The primary purpose of this patent framework was to provide an NFT-based concept that could be used to patent a decentralized, anti-tamper, and reliable network for trade and exchange around the world. Finally, we addressed several open challenges to NFT-based inventions.

### Acknowledgements
This work has been partially supported by CAS President's International Fellowship Initiative, China [grant number 2021VTB0002, 2021] and National Natural Science Foundation of China (No. 61902385).

### Author contributions
S.M. Hosseini Bamakan: Conceptualization, Methodology, Supervision, Project administration, Writing—review and editing Nasim Nezhadsistani: Formal analysis, Data curation, Visualization, Writing—Original Draft Omid Bodaghi: Formal analysis, Resources, Visualization, Writing—Review and editing Qiang Qu: Supervision, Project administration.

### Competing interests
The authors declare no competing interests.

### Additional information
**Correspondence** and requests for materials should be addressed to Q.Q.

**Reprints and permissions information** is available at www.nature.com/reprints.

**Publisher's note** Springer Nature remains neutral with regard to jurisdictional claims in published maps and institutional affiliations.